# Estimands in Hematologic Oncology Trials

Steven Sun (Janssen Research and Development LLC)
Hans-Jochen Weber (Novartis Oncology Schweiz AG)
Emily Butler (GlaxoSmithKline)
Kaspar Rufibach (F. Hoffmann-La Roche AG)
Satrajit Roychoudhury (Pfizer)

**Abstract**

The estimand framework included in the addendum to the ICH E9 guideline facilitates discussions to ensure alignment between the key question of interest, the analysis, and interpretation. Therapeutic knowledge and drug mechanism play a crucial role in determining the strategy and defining the estimand for clinical trial designs. Clinical trials in patients with hematological malignancies often present unique challenges for trial design due to complexity of treatment options and existence of potential curative but highly risky procedures, e.g. stem cell transplant or treatment sequence across different phases (induction, consolidation, maintenance). Here, we illustrate how to apply the estimand framework in hematological clinical trials and how the estimand framework can address potential difficulties in trial result interpretation.

This paper is a result of a cross-industry collaboration to connect the International Conference on Harmonisation (ICH) E9 addendum concepts to applications. Three randomized phase 3 trials will be used to consider common challenges including intercurrent events in hematologic oncology trials to illustrate different scientific questions and the consequences of the estimand choice for trial design, data collection, analysis, and interpretation. Template language for describing estimand in both study protocols and statistical analysis plans is suggested for statisticians' reference.

**Key Words:** Estimand, hematology clinical trials, Intercurrent events, ICH E9 addendum

**Section 1: Introduction**

The key questions of interest in a clinical trial vary from one stakeholder to another. The current practice of defining trial objectives rather vaguely has led to various challenges in interactions with different stakeholders, including regulators and payers. To improve transparency and ensure alignment between trials objectives, data collection and statistical approaches, it appears necessary to go back to the first principles and to clearly define what is to be estimated.

The addendum to the ICH E9 guideline was released in 2019 to introduce the estimand framework. This framework targets to align the clinical trial objectives with the design and statistical analyses by providing detailed definitions of the quantity of scientific interest

("estimand"). Besides estimands, the addendum also discusses the role of sensitivity and supplementary analyses. Therapeutic knowledge and drug mechanism play a crucial role in determining the estimand of interest and the design (Kenneth 2020).

In recent years, many clinical trials have been initiated in hematologic malignancies due to an enhanced understanding of cancer pathologies (Palmisiano 2018). Hematological malignancies such as leukemia, multiple myeloma and lymphoma represent a subset of indications in hematologic oncology (referred to as "hematology" for simplicity below). Clinical trials in hematologic malignancies are characterized by specific definition of endpoints, data collection, magnitude of clinically meaningful treatment effect, and interpretation of treatment effect. Moreover, standard therapies are often following a sequence of treatment (e.g. induction, consolidation, maintenance) and offering stem cell transplantation can be an important part of such a treatment strategy. This can lead to challenges when defining the treatment attribute of an estimand (Roussel 2019, Salwender 2019). Time-to-event (TTE) endpoints are mainstays of late phase hematology clinical trials. Though, overall survival (OS) is considered the gold standard for determining treatment efficacy in oncology trials, but OS can be challenging to assess because of long trial durations and the potentially confounding impact of subsequent therapies. Therefore, other clinical endpoints like progression-free survival (PFS), event-free survival (EFS), time to progression and response rate (RR) may be appropriate endpoints for the regulatory approval of treatments (Ellis 2014, Johnson 2003 and Smith 2017). Some statistical analyses may be more intuitive and easier to interpret than others, depending on the key question of interest, trial type, patient population, and duration of follow-up. Another typical feature for a hematological malignancy is that the broad population defined in a clinical trial for the same type of cancer is often heterogeneous in the sense that patients could have very different types of genomic characteristics. For example, some patients with diffuse large B cell lymphoma (DLBCL) may be curable while others may die quickly irrespective of any treatment options available so far (Zhang 2013).

In this paper, we present the utility of the estimand framework for typical hematology trials and provide practical recommendations. The estimand framework lends itself to a more transparent way of specifying each objective of a trial and ensuring alignment with the selected estimator/analysis method so that the most appropriate methods for primary and sensitivity analyses are identified and applied. In this paper we discuss ways to incorporate the estimand framework, especially strategies to common intercurrent events when designing a clinical trial for hematology malignancies. Issues relevant to defining estimands are discussed. On the side of the estimator, considerations are given regarding the trial design to minimize dropouts and missing values. In addition, we present real examples to illustrate the application of different estimand strategies.

The rest of the paper is structured as follows. In Section 2, we introduce three typical case trials to illustrate the key aspects of the estimand framework for hematology trials. Section 3 and 4 focus on the strategies for different intercurrent events and relevant sensitivity analyses and supplementary analyses. Finally, the impact of the estimand framework on trial design, data collection and data analysis are discussed in Section 5. Section 6 concludes with a discussion. Sample language for protocol and statistical analysis plan (SAP) are provided in the Appendix.

**Section 2: Motivating examples**

We provide three case studies to demonstrate the utility of the estimand framework in confirmatory hematology trials. These examples consist of randomized Phase 3 trials with time to event endpoint comparing a novel treatment with standard of care (SOC). They include the disease areas non-Hodgkin's lymphoma, multiple myeloma, and acute myeloid leukemia. The purpose of this exercise is to demonstrate the application of the estimand framework in hematology to practitioners. We have no intention to comment or judge the clinical activity of the drugs involved or related regulatory decisions. Further details of the treatment, population, primary endpoint, and design are shown in Table 2 and will be used in later sections for introducing appropriate estimands.

**Table 1  Design Specification of Three Examples**

| Examples | **Gallium Study** [Marcus 2017] | **Multiple Myeloma Study** [a] | **RATIFY Study** [Stone 2017] |
|---|---|---|---|
| Treatment | Open label randomized study comparing<br>• Obinutuzumab<br>• Rituximab<br>in combination with three backbone chemotherapies | Open label randomized study comparing<br>• Drug X + background therapy<br>• background therapy | Double-blinded study<br><br>• Midostaurin<br>• Placebo<br>in combination with chemotherapy |
| Population | Patients with untreated advanced indolent non-Hodgkin's lymphoma | Patients newly diagnosed with multiple myeloma and eligible for high-dose therapy | Patients with first-line acute myeloid leukemia (AML) with a FLT-3 mutation |
| Design | See Figure 1a | See Figure 1b | See Figure 1c |
| Induction Phase | Patients have received Obinutuzumab or Rituximab + chemotherapy | Patients have received drug X + background therapy or background therapy only. Induction phase is followed by a consolidation phase. Induction and consolidation phase will consist of six 28-day cycles, 4 cycles | Patients have received midostaurin or placebo along with chemotherapy for one cycle. If there are definitive evidence of clinically significant residual leukemia, a second cycle of same therapy continues. Induction phase is |
| Maintenance Phase | Participants achieving response (Complete response [CR] or partial response [PR]) continue the treatment with the antibody only until disease progression. Those with stable disease (SD) undergo the same schedule of assessments, but do not receive further therapy. | of induction, followed by ASCT, then 2 cycles of consolidation<br><br>Patients are followed by background maintenance therapy with or without drug X until disease progression or unacceptable toxicity | followed by 4 cycles of consolidation<br><br>Patients who remained in remission entered a maintenance phase in which they received drug A or placebo. |
| Primary Endpoint | Progression-free survival | Progression-free survival | Overall survival |
| Sample Size (events) | 1202 | 690 | 717 |

[a]: This study is still ongoing

Further details of the three trial designs are schematically shown in Figure 1. For all three studies, the design has two phases: induction and maintenance. In addition to that the multiple myeloma and the RATIFY trial also have a consolidation phase. During the induction (or consolidation) phase, patients are treated aggressively with the experimental drug and/or SOC. This is followed by maintenance therapy until disease progression or unacceptable toxicity. For all the trials, the comparisons are made with a primary endpoint which is measured over all phases following a traditional intent to treat (ITT) principle. However, the design cannot address other important clinical questions such as the benefit of treatment in each phase due to the occurrence of different intercurrent events (e.g., start of maintenance therapy, stem cell transplantation (ASCT or HSCT), possible response or remission which decides the maintenance therapy etc.). Key assumption like proportional hazard between two treatment groups is questionable in presence of these intercurrent events (e.g., ASCT may change the disease course completely). Other challenges include interest in additional patient populations (marginal-zone lymphoma (MZL) in Gallium trial, for which the primary interest is in follicular lymphoma (FL) patients), discrepancies between investigator and independent reviewer regarding the assessment of progression, and assessment of secondary endpoints (e.g., event-free survival, EFS). In the next section we show how the five estimand attributes can be used to address these issues and answer appropriate clinical questions of interest.

**Figure 1:** Schematic Overview of the Study Designs for Three Examples

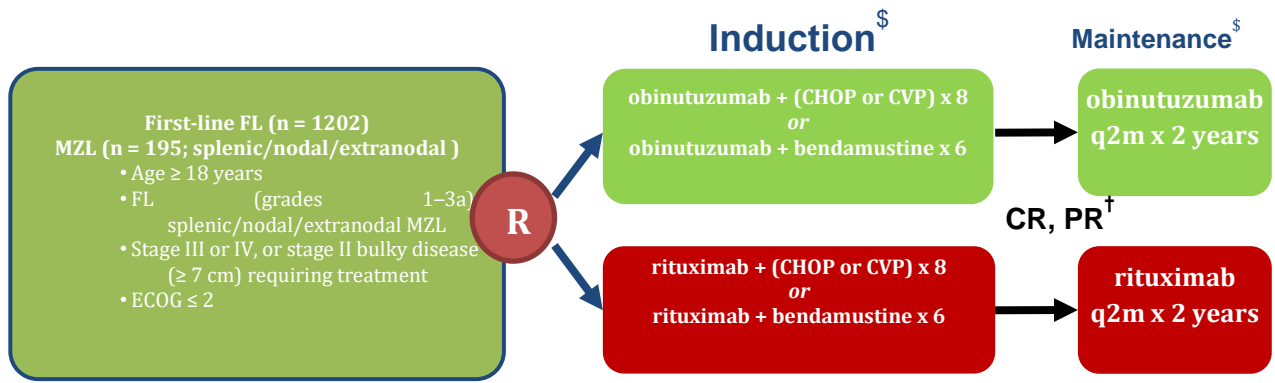

1a: Gallium Study

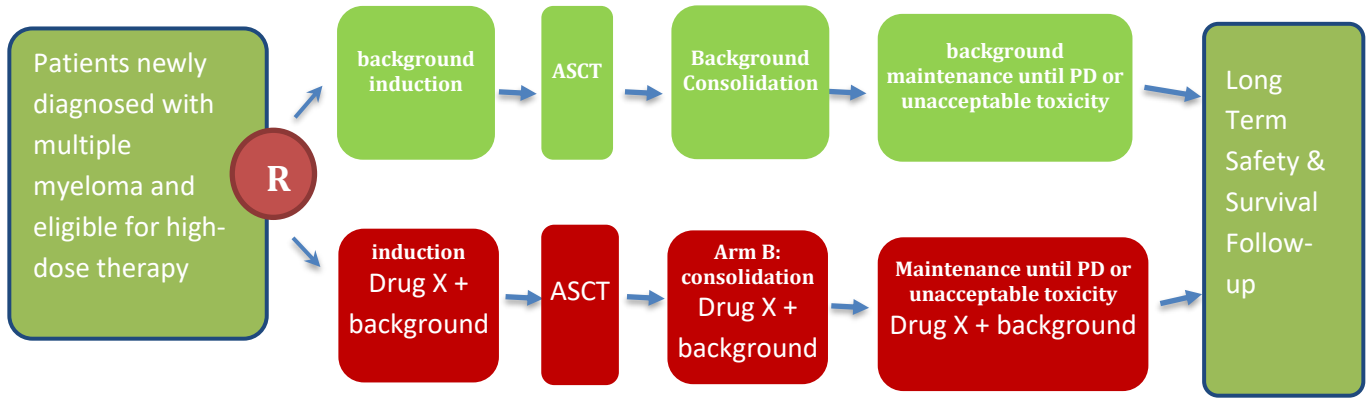

1b: Multiple Myeloma Study

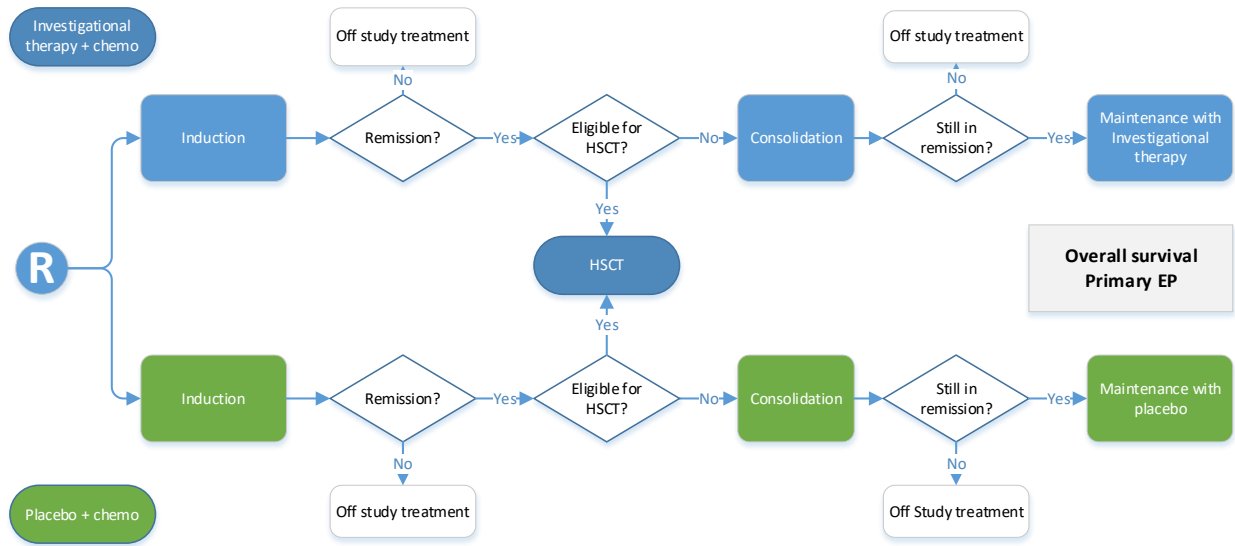

**1c: RATIFY Study**

For all the examples above, intent-to-treat (ITT) analyses were planned for the primary analysis. The clinical meaningfulness of such estimates needs to be assessed as this does not consider that some patients take new anti-cancer therapy during the trial. This may convolute the overall biological effect of the treatment. For example, consider a situation where most patients on placebo switch to a new anti-cancer therapy but only a few on the active drug switch. In this case, the ITT analysis essentially compares the new drug with another anti-cancer therapy, rather than with placebo. Moreover, a series of sensitivity analysis are often pre-specified in the analysis plan to show robustness of the treatment effect. Often, these analyses lack the description of the underlying estimand which creates issues with clinical interpretation. In the next section we show the utility of the estimand framework to address these challenges.

**Section 3: Application of the estimand framework in hematologic oncology**

In this section we discuss how the key questions in hematological studies can be viewed in the estimand framework. Based on the ICH E9 addendum it is defined through the five attributes viz, treatment, population, variable, intercurrent event, and population-level summary. Each intercurrent event and associated strategy needs to be clearly stated. Intercurrent events are clinical events that occur after treatment initiation that affect either the interpretation or the existence of the measurements associated with the clinical question of interest . Some potential examples of intercurrent events in hematology trials are discontinuation of treatment, start of maintenance therapy, or treatment switching. The treatment effect is quantified using a population-level summary. There could be multiple ways of describing the same estimand. For example, a composite strategy for an intercurrent event of subsequent therapy  together with the endpoint of PFS is the same as treatment policy strategy for an intercurrent event of subsequent therapy when the endpoint is defined as EFS, which is time interval from randomization to the earliest of PD, subsequently therapy or death.

All these case studies in Section 2 are large phase 3, randomized, controlled trials in patient populations with different hematologic indications and represent typical trial designs in hematology drug development. For each example, we now describe the primary scientific question and the corresponding five estimand attributes. These examples help practitioners to understand how to explain the key question of interest using the estimand framework in the study protocol and analysis plan. All the details are provided in Table 2. Note that these trials were designed and executed prior to the release of the ICH E9 addendum. Therefore, this exercise is performed retrospectively for illustration purposes only.

The case studies represent typical treatment approaches in hematologic indications consisting of different treatment parts. The experimental treatment is often added to the standard of care treatment. Trial designs comparing treatment strategies (combination or treatment sequence) will allow proper assessment of the whole strategy. However, it is extremely difficult if not impossible to estimate the contribution of an individual component of the strategy unless $2^{nd}$ randomization or multiple trials are implemented. Health authorities might request such information if some parts of the treatment strategy are not well established like the maintenance treatment in AML or myeloma. With the estmand framework, there is no ambiguity about what treatment effect the study is designed to establish.

The same considerations apply to stem cell transplantation (SCT). In many hematologic diseases, SCT is part of the standard treatment options. In AML, SCT improves the chance of cure but relates also to specific risks like graft-vs-host disease or transplant-related mortality. As such, SCT might impact the results for OS or PFS. The question is whether we need to assess the impact of SCT on OS or PFS. However, the receiving SCT might be a result of the experimental therapy and thus, assessing the contribution of the elements of a treatment strategy might be challenging or even not be feasible. Assessing the effect of the whole treatment strategy integrating all its components might be the most relevant approach for the patients. The estimand framework will facilitate these discussions with the health authorities prior to starting a study. More specifically, sponsor and the health authority can agree on what can be estimated and what cannot be estimated at design stage.

**Table 2: Attributes of Estimands as per ICH E9 Addendum for Three Case Studies**

|  | **Gallium study** | **Multiple myeloma study** | **RATIFY study** |
|---|---|---|---|
| **Scientific question** | Will the addition of Obinutuzumab to FL patients treatment strategy prolong the time to death and progression regardless of new anti-lymphoma treatments prior to experiencing a PFS event | Will the addition of new drug to VRd to a treatment strategy consisting of induction, ASCT, consolidation and maintenance prolong the time to death and progression compared to VRd alone if patients with new therapy would have had the same subsequent event hazards as patients without new therapy, had they not taken new therapy | Will the addition of Midostaurin to newly diagnosed AML treatment strategy prolong the time to death regardless of new therapies and SCT |
| **Primary Objective** | To evaluate the efficacy of G-chemo followed by G-maintenance therapy compared with R-chemo followed by R-maintenance therapy in patients with previously untreated advanced FL, as measured by investigator-assessed PFS. | To evaluate the efficacy of the new drug to a standard care VRd as a treatment sequence in patients with newly diagnosed multiple myeloma who are transplant eligible in terms of PFS using algorithm. | To evaluate benefit of Midostaurin in addition to the standard care in prolonging overall survival compared with the standard care in patients newly diagnosed AML treatment strategy irrespective of any supportive care and subsequent therapy use. |
| Treatment | Patients receive 6 or 8 induction cycles of induction chemoimmunotherapy. Responding patients receive single agent therapy during maintenance therapy for up to 2 years | Patients receive 4 cycles induction therapy. Responding patients receive ASCT followed by 2 cycles of consolidation therapy. Thereafter patients are entering maintenance therapy | Patients receive 1 cycle induction therapy. If no remission, 2$^{nd}$ induction cycle is given. Eligible patients receive SCT. Otherwise up to 4 cycles of consolidation. After consolidation patients receive 12 cycles single-agent maintenance treatment |

| Population | FL patients as defined by protocol eligibility criteria | MM patients as defined by protocol eligibility criteria | First-line AML patients as defined by eligibility criteria |
|---|---|---|---|
| Variable | PFS defined as time from randomization to PD as per investigator assessment or death, whichever is earlier | PFS by algorithm based on lab data and clinical data | OS |
| Intercurrent events and strategies | | | |
| 1: New anti-cancer therapy prior to PFS event | Treatment Policy: intercurrent event is ignored#, PFS time irrespective of start of new therapies is considered | Hypothetical: Patients who received anti-cancer therapy are assumed to have the same risk as those who did not receive subsequent anti-cancer therapy | Treatment Policy: intercurrent event is ignored, entire survival time after starting new therapies is taken into account |
| 2: Treatment discontinuation | Treatment Policy: intercurrent event is ignored# | Treatment Policy: intercurrent event is ignored | Treatment Policy: intercurrent event is ignored |
| Population-level summary | Hazard ratio* | Hazard ratio* | Hazard ratio* |

#: The occurrence of the intercurrent event is considered irrelevant in defining the treatment effect of interest
*Hazard ratio calculated using stratified cox regression with treatment as the only covariate

The estimands described in Table 2 clearly define what is to be estimated and how to estimate treatment benefit based on the scientific questions of interest. The estimand framework ensures consistency of the choice of statistical analysis with the scientific question of interest (Holzhauer 2015). This avoids misalignment between the trial objectives, analysis, and interpretation of result (Akacha 2017). Different strategies used to address intercurrent event of subsequent therapy address different clinical questions and this becomes clearer when embedding these clinical questions in the estimand framework. Different stakeholders will likely be interested in different scientific questions. For example, health authorities may be interested in the OS benefit irrespective of subsequent anti-cancer therapy while HTA (Health Technology Assessment) is interested in the OS benefit if no subsequent anticancer is allowed.

Under the estimand framework, ambiguity of interpretation of trial results will be eliminated or minimized. Each scientific question of importance can be addressed by the mutual agreement of estimand description between trial sponsor and the targeted stakeholder (health authorities, HTA, patients, etc.) for assessing the trial results.

**Section 4: Sensitivity analysis vs Supplementary analysis -**

Description of estimand as illustrated in case studies in Section 3 above includes the main estimator or summary of treatment effect. The ICH E9 addendum also provides guideline to assess the robustness of treatment effect. An estimand is associated with one specific scientific question of interest. However, there could be multiple scientific questions that need to be answered from a clinical trial to establish the benefit of a treatment intervention. Consequently, there are two types of analyses that are important in understanding the impact of treatment intervention.

Sensitivity analyses are a series of analyses targeting the same estimand. This means there is usually no change to the treatment, population, variable, handling of intercurrent events, and population-level summary measure. These analyses assess different assumptions of the statistical methods to explore the robustness of the statistical inference, such as the main estimator and deviations from its underlying modelling assumptions or limitations in the data (Mehrotra 2016). Note that this redefines the term "sensitivity analyses" compared to its traditional pre-addendum use.

Supplementary analyses target a different estimands of interest. This section focuses on sensitivity and supplementary analysis for PFS and OS, illustrated by the case studies introduced in Section 2. Please note the recommendation below is linked to the GALLIUM study only. In general, whether analyses are sensitivity or supplemental, or appropriate at all, is dependent on the particular context and strategy. Different analyses may serve as the primary analyses for different stakeholders. A supplementary analysis for one stakeholder or in particular context may be the primary or a sensitivity analysis for another stakeholder or in another context.

**Section 4.1 Progression Free Survival**

In this section, we provide the working group's recommendations how to embed these analyses for PFS in the addendum's framework of primary, sensitivity, and supplementary analyses. Progression free survival is defined as the time from randomization until death from any cause or objective disease progression. PFS is an acceptable endpoint for oncology treatment approval in many tumor types (FDA guidance 2018) . Table 3 summarizes common analyses performed for PFS using Gallium study for illustration purpose in addition to the main analysis as stated in Section 3. The primary analysis is a stratified Cox regression analysis based on investigators' assessed PFS irrespective of any subsequent anti-cancer therapy use. The main analysis is aligned with the scientific question presented in Table 2.The concept described in here applies also to related endpoint definitions like event-free survival (EFS) in AML.

**Table 3: Common analyses performed for endpoint PFS (GALLIUM study)**

| Analysis method | Key assumptions | Sensitivity or supplementary | Rationale | Recommendation |
|---|---|---|---|---|
| Same analysis as main analysis except for using PFS by independent review committee (IRC-PFS) | IRC-PFS is perceived as less biased compared with INV-PFS since this is an open-label study. As in the main analysis, patients are assumed to be consistently followed beyond subsequent therapy | Sensitivity analysis | Analysis corresponds to the same primary estimand as stated in the Table 2. It is considered as data limitation for disease assessment | Same disease assessment criteria should be used for both investigators' assessment and IRC assessment. IRC assessment may introduce bias due to informative censoring |
| Unstratified Cox regression | Common baseline risk among different strata | Sensitivity analysis | Same estimand as in Table 2. The main analysis assumes different baseline risk across different strata. This analysis is based on an assumption which deviates from the assumption used in the main analysis | Useful analysis for checking robustness of the main analysis |
| Same analysis as the main analysis except for imputing the event time of subjects considered to have problematic informative censoring [Liu 2017] | Informative censoring | Sensitivity analysis | To demonstrate the impact of violations in necessary assumptions, and assesses robustness of the p-value as calculated from imputed data as compared with un-imputed data | Checking if the distribution of censoring was approximately random or balanced between the arms |
| Censoring for 2 or more consecutive | Non-informative missing | Sensitivity analysis | Targets the same estimand. It is considered as data limitation and | In some cases, if the consecutive missing is due to an intercurrent |
| missing assessment(s) | | | missing assessment is assumed to be non-informative | event such as health care facility closure temporally due to pandemic, then it can be viewed as supplementary analysis since it corresponds to a different estimand (hypothetical, see Degtyarev 2020) |
| Analysis using per-protocol (PP) population | PP population represents a subset of patients, which adhere to the protocol defined condition. | Supplementary analysis | Addresses the population attribute of estimand. PP population definition often depends on the outcome of treatment, such as completing a minimum of 6 cycles of treatment. | It is highly recommended not to have any such analysis. Patients in PP population do not represent random samples of the target population as defined by the eligibility criteria. Consequently, the interpretation is very difficult and such analysis is not useful. |
| Censoring at subsequent therapy | Subsequent therapy impacts the interpretation of treatment effect | Supplementary analysis | Different strategy for intercurrent and hence it corresponds to different estimand. It is a hypothetical strategy and estimation via simple censoring assumes non-informative censoring at subsequent therapy use | Useful analyses to understand the treatment effect from different perspectives. In the past, the FDA could accept this as the main analysis while the EMA often preferred treatment policy strategy for the intercurrent event of subsequent therapy |
| covariate-adjusted analysis (multi-variate Cox regression analysis) [Permutt 2020] | Covariates may be confounding factors for disease outcome and imbalance may exist between two treatment arms | Supplementary analysis | The analysis addresses a different scientific question. i.e., what is the treatment benefit of Obinutuzumab in terms of PFS for patients who have covariates valued at population means. Covariate adjusted multivariate Cox regression provides a conditional treatment effect rather than marginal treatment effect. | Interpretation should be careful, and results may not be consistent since it does not address the same estimand |

**Section 4.2 Overall Survival**

Now, we provide some recommendations how to embed addendum's framework of primary, sensitivity, and supplementary analyses for Overall Survival (OS). Overall survival is defined as the time from randomization until death from any cause. Survival is considered the most reliable cancer endpoint, and when trials can be conducted to adequately assess survival, it is the preferred endpoint ( FDA Guidance 2018, EMA/CHMP Guidance 2017). OS is often performed based on survival data collected in the study for all patients. Table 4 lists common analyses performed for OS in addition to the main analysis as stated in Section 3, which is the stratified Cox regression analysis using the survival data collected in the entire study duration ignoring any subsequent anti-cancer therapy use.

**Table 4: Common analyses performed for endpoint OS**

| Analysis method | Key assumptions | Sensitivity or supplementary | Rationale | Recommendation |
|---|---|---|---|---|
| Censor at subsequent anti-cancer therapy | Patients who received subsequent anti-cancer therapy have the same risk as those who did not receive | Supplementary analysis | Using the hypothetical strategy to account for the intercurrent event, new anticancer therapy, which is a new estimand. | Not very useful since it is unlikely that assumption holds |
| | subsequent anti-cancer therapy | | | |
| Rank preserving structural failure time (RPSFT) model | Treatment effect is the same regardless of when the experimental treatment is initiated | Supplementary analysis | The analysis is to address a different question: what is the treatment benefit if patients in the Rituximab arm are not allowed to receive Obinutuzumab after PD. The treatment effect is expressed as an acceleration factor through G-estimation or by estimating hazard ratios using reconstructed data through the Cox model. Different strategy (an alternative hypothetical strategy) for intercurrent of treatment crossover (control to experiment arm). Hence it corresponds to different estimand. | Useful analysis to adjust for crossover. This method is advantageous as it preserves randomization, does not require information on covariates and can handle larger proportion of patient switching. This analysis is often of interest to the HTA for reimbursement decision |
| Inverse Probability Censoring Weighting (IPCW) analysis | All common predictors for survival and/or the probability of treatment switching are measured | Supplementary analysis | The corresponding scientific question is "what would be the treatment difference if subsequent anti-cancer therapies are not available?" It is one of hypothetical strategies used for intercurrent event of subsequent anti-cancer therapy. | Clinical interpretation is very different from the main OS analysis. However, it provides some insight in understanding the impact of treatment intervention. This analysis is often of interest to the HTA for |

| | | | | reimbursement decision |
|---|---|---|---|---|
| Two-stage approach (Latimer 2014, Alshurafa 2012) | OS benefit adjusted for treatment crossover at a specific disease-related time-point progression (PD). It requires not only the complete data collection at PD, but also the assumption that switching occurs soon after PD | Supplementary analysis | The treatment effect for those patients who cross over is removed, i.e. survival time after progression is adjusted using a hypothetical strategy. The treatment effects are estimated separately for crossover patients from control group and for patients originally randomized to the investigational drug group using an accelerated failure time (AFT) model | The two-stage approach assumes no unmeasured confounders. This assumption can be evaluated by running a full and reduced model for comparison. |

**Section 5: Impact on trial design, data collection, and data analysis**

The design of a clinical trial needs to be aligned with the choice of estimand or estimands which reflects the primary objectives of the clinical trial. The agreed estimands dictate the data that needs to be collected. Different estimands might have different requirement of data collection. More importantly, a clinical trial may have multiple estimands, which implies that the data collection should be determined by the need to address them all (Scharfstein 2019, Section A.4 of ICH E9 (R1) 2019).

For Gallium study, the study objective was to evaluate the efficacy of G-chemo followed by G-maintenance therapy compared with R-chemo followed by R-maintenance therapy in patients with previously untreated advanced FL, as measured by investigator-assessed PFS. The primary estimand was the hazard ratio of G-chemo followed by G-maintenance therapy over R-chemo followed by R-maintenance therapy in the target population, regardless of the subsequent anti-lymphoma therapy use or whether patients discontinued planed treatment. The study design was aligned with the primary objective. If there is another objective to investigate if G-maintenance is more efficacious than R-maintenance, then at the end of induction phase, for those who had PR or better, they should be re-randomized into G-maintenance or R-maintenance. Similarly, for the 2[nd] and 3[rd] case study in patients with multiple myeloma and AML respectively, the study objective was to evaluate the efficacy of entire treatment sequence (induction followed by consolidation and then maintenance), the design described in Section 2 was the most efficient one. However, the design will not allow the causal inference of added effect of the new treatment in the maintenance. A potential challenge with this design is that the observed PFS benefit may be driven by the benefit of new drug in induction/consolidation only. A $2^{nd}$ randomization after induction/consolidation would be needed if it is of interest to characterize the maintenance effect.

In both Gallium and RATIFY studies, the strategy for intercurrent events such as subsequent therapy and treatment discontinuation is as treatment policy. For the multiple myeloma case study, even though the primary estimand used hypothetical strategy for the intercurrent event of subsequent therapy, a supplementary analysis corresponding to treatment policy strategy for subsequent therapy was planned. Therefore, data for disease assessment should be collected until patients progress, or die if no progression or study end, whichever is earlier. The focus of discussion has been on the estimand with PFS/OS endpoints so far. In fact, PFS2, which is defined from randomization to progression corresponding to the first subsequent therapy or death, is also of interest from regulatory perspective in most of cases. Consequently, disease assessment, especially PD assessment on the next line subsequent therapy should also be collected.

From data analysis perspective, an analytic approach, or estimator, should be aligned with the given estimand and should be able to provide an estimate on which a reliable interpretation can be based. This implies that any assumptions should be explicitly stated, and sensitivity analyses should be carefully described to assess the robustness of the results to the underlying assumptions. In Gallium study, analyses performed on PFS included different population (FL only and overall population), different censoring schemes, and with or with adjustment for stratification and/or baseline covariate (Rufibach 2019). Some of them were targeted on the same estimand and should be clarified as sensitivity analysis but others should be specified as supplementary analysis to avoid difficulty in interpretation. The subtlety between supplementary and sensitivity analysis is illustrated in Section 4. For the $2^{nd}$ case study in patients with multiple myeloma, we may not be able to characterize the benefit magnitude of the new drug when added into the lenalidomide for the maintenance for the given study design, however, it is still of interest to demonstrate adding the new drug to lenalidomide during the maintenance does contribute to the overall benefit as the entire treatment sequence. Therefore, supplementary analyses such as treating maintenance as the time dependent covariate or new PFS with reference time as the start of maintenance might be to be included in the analysis plan. The treatment regimen in the $3^{rd}$ case study in patients with AML is even more complicated. Like the other 2 trials, it has multiple treatment phases. In addition, transplantation was performed at the discretion of the investigator. The contribution of transplantation to the treatment strategy may be of interest. Supplementary analysis to tease out the effect of transplantation should be included in the analysis plan even if transplantation itself is part of the treatment regimen.

**Section 6: Conclusion and Discussion**
The estimand framework is an efficient tool to ensure consistency between the scientific question and the definition of the study objectives. Per E9(R1), intercurrent events are not to be thought of as a drawback to be avoided. Discontinuation of treatment, changing treatment, etc. are part of clinical practice and are part of clinical trials. The framework ensures transparency in unfavorable yet unavoidable situations in clinical trials ("intercurrent events"). The outcome of studies that are following the estimand framework can be interpreted in a consistent manner

(Bengoudifa 2018). We highly recommend estimand(s) to be included in the study protocol and details including sensitivity analyses and supplementary analyses to be included in the statistical analysis plan (SAP) for the study. Some sensitivity analysis approaches such as tipping point analysis recommended in the estimand guidance document but not elaborated in this paper may also be considered in the SAP. Recommended template language can be found in the appendix. The 3 case studies discussed in this paper represent typical treatment regimens and study designs for drug development in hematologic oncology. They share a common feature of having multiple treatment phases and the treatment in the follow-up phase may depend on the treatment effect on the previous phase. One potential issue with multiphase treatment is the violation of proportional hazard assumption for PFS analysis. Assumption of proportional hazard should be checked whenever Cox regression model is used. For non-proportional HR, the traditional Cox regression model to characterize the treatment benefit using HR may be difficult to interpret (Byod 2012, Gregson 2019). Other analysis methods with different population level summary such as restricted mean survival time (RMST) analysis and responder analysis using pre-specified milestone PFS/OS proportion may need to be adapted. Common intercurrent events and potential strategies are discussed in detail for each case study. But the choice of strategies should depend on the disease as well as the mechanism of action of the drug (Leuchs 2015, 2017). We only illustrated one estimand corresponding to the primary objective in each case study. However, one estimand may not be enough for the study objective. For example, in the $2^{nd}$ case study, characterizing the benefit of the new drug when adding in the standard care regimen in induction phase, induction/consolidation phase are important together with the overall benefit for the entire treatment sequence with long term follow-up are equally important. Especially from regulatory perspective, the observed overall treatment benefit is not driven by one single phase. The formulation of the scientific question of interest requires an a priori discussion amongst stakeholders to tailor the estimand definition and the treatment strategy attribute accordingly: To assess the contribution of each treatment phase of a sequential treatment strategy might require re-randomization of eligible patients when starting a subsequent treatment phase. This might operationally be not feasible in many hematologic malignancies with rare patient populations.

Another challenge introduced in the Gallium study is the multiplicity control. It is not uncommon to include multiple tumor types of the same pathology in one study (both FL and MZL in Gallium study). Analysis on different population corresponds to different estimands. So, the significance level for the analysis associated with one estimand must be considered in the context of the analysis associated with other estimands. In essence, the estimand framework is about an estimation (what to estimate and how to estimate). Hypothesis testing is important for decision making but it should be handled beyond the estimand framework even though the same analytical approach may be used for both hypothesis testing and estimation of treatment effect.

Recently, CAR-T therapies have shown promising results for multiple hematological cancer types. The development of CAR-T therapies poses unique challenges from statistical analysis perspective. E.g., bridging treatment is usually allowed during the manufacturing period and conditioning lymphodepletion chemotherapy is required before infusion of CAR-T cells

(Degtyarev 2019). For single arm study, should the response rate be provided for all patients who received CAR-T infusion or for all patients who had apheresis? Similarly, what is the appropriate way for measuring treatment benefit for randomized control trials. There won't be any ambiguity for the treatment effect associated with the defined 'treatment' (strategy a: sequence of bridging treatment, conditioning therapy and infusion of CAR-T vs strategy b: infusion of CAR-T) under estimand framework. Emphasis of this paper is placed on the recommendation of description of estimands and careful selection of sensitivity analyses and supplementary analyses for hematological trials. Data collection and analysis should also be aligned in coherent manner to avoid disconnect between trial objectives and estimands. The template language on estimands for protocols and SAPs in the appendix can serve as detailed instructions/ recommendations.

APPENDIX

Sample template protocol and statistical analysis plan language (SAP) are provided below using GALLIUM study for illustration purpose. Please note that the language below is related to estimand only. Hypothesis testing should be included in a separate section independent of estimand description since estimand framework focus on the estimation of treatment effect only.

**Protocol Language**

*Italic* text is guidance and should be deleted from the final version.

X. Objectives and Endpoints

*Note: Estimands are not required for all secondary or exploratory endpoints. Only estimands for critical secondary endpoints that pertain to regulatory decision-making should be included. If needed, the language for secondary endpoint should mimic the primary estimand*

| Objective | Primary Estimand | | | | |
|---|---|---|---|---|---|
| | Population | Variable | Summary Measure | Treatment | potential intercurrent events |
| The primary trial objective is to demonstrate superiority of the experimental over the control treatment. The primary comparison of progression-free survival (PFS) will be made regardless of whether patients withdraw from treatment or receive new-anti lymphoma therapy prior to disease progression | FL patients, defined by list of in- and exclusion criteria | Investigator assessed PFS, defined from rando to PD or death | Hazard ratio based on Cox regression model stratified by chemotherapy and FLIPI1 | G added to chemotherapy followed by G-maintenance therapy vs R added to chemotherapy followed by R maintenance therapy. SCT as option. | (1) New anti-lymphoma therapy (NALT) (2) treatment discontinuation |

**Statistical Analysis Plan Language**

*Italic* text is guidance and should be deleted from the final version.

X. Main analytical approach for the primary estimand

The primary estimand corresponding to the primary endpoint is defined as:

1. Treatment: 6 or 8 21-day cycles obinutuzumab D1 + C1D8, C1D15: 1000mg/m2 flat + site-specific choice of CT (CVP, Benda, CHOP) in induction followed in responding patients by 1000mg flat every 2 months until PD or up to 2y with 6 or 8 21-day cycles rituximab 375mg/m2 D1 + site-specific choice of CT (CVP, Benda, CHOP) in induction followed in responding patients by 375mg/m2 every 2 months until PD or up to 2y
2. Population: first-line follicular lymphoma patients as defined by the inclusion and exclusion criteria;
3. Variable: PFS (time from randomization to the earlier of disease progression or death)
4. Population level summary: hazard ratio

Intercurrent events under consideration: 1)NALT, 2) treatment discontinuation

NALT and treatment discontinuation will all be ignored (treatment policy strategy).

The main analysis will be based on investigators' assessed PFS. Stratified Cox regression model with stratification factors chemotherapy and FLIPI1 will be used and treatment is included as the only independent variable in the regression model. The model assumption is proportional hazard ratio with each stratum, but the baseline risk may be different for different stratum. Patients will be censored at the last disease assessment if no PFS event is observed during the study (administrative censoring).

X. Sensitivity analysis for the primary endpoint

There are two planned sensitivity analyses that both target the primary estimand. The first will be to relax the assumption that chemotherapy and FLIPI1 are confounding factors, so the analysis will be repeated but unstratified, which implies the baseline risk is the same across strata The other sensitivity analysis is based on IRC assessed PFS and the same analysis as the main analysis will be performed. This is to address the data limitation related to the assessment of the primary endpoint.

X. Supplementary analysis for the primary endpoint

There are three planned supplementary analyses. These are considered supplementary analyses as they inherently change the estimand. Per the ICH E9 addendum recommendations, only one thing per analysis is changed at a time

The first supplementary analysis changes which strategy is employed for the intercurrent event of subjects who withdrawal. Withdrawal of obinutuzumab will be considered part of PFS definition and will be considered an event (composite strategy) and withdrawal of rituximab will be censored (hypothetical strategy)

The second supplementary analysis changes which strategy is employed for the intercurrent event of NALT. If a subject took a NALT they were censored (hypothetical strategy).

The third supplementary analysis changes which strategy is employed for the intercurrent event of discontinuation of study treatment. If a subject discontinues study treatment prior to PD or clinical cutoff, it is considered an event as part of PFS (composite strategy).

ACKNOWLEDGEMENTS

We would like to thank our cross-industry working group member Marie-Laure Casadebaig from Celgene, Viktoriya Stalbovskaya from Merus, and Evgeny Degtyarev from Novartis for their inputs and comments.